# Photoluminescence from single walled carbon nanotubes: a comparison between suspended and micelle-encapsulated nanotubes

J. Lefebvre[1], J. M. Fraser[1], Y. Homma[2], P. Finnie[1]

[1]Institute for Microstructural Sciences, National Research Council, Building M-50, Montreal Road, Ottawa, Ontario, K1A OR6, Canada.

[2]NTT Basic Research Laboratories, Nippon Telegraph and Telephone Corporation, 3-1 Morinosato-Wakamiya, Atsugi, Kanagawa 243-0198, Japan

**Abstract.** Single walled carbon nanotubes (SWNTs) are luminescent. Up to now, two preparation methods, both of which isolate individual SWNTs, have enabled the detection of nanotube bandgap photoluminescence (PL): encapsulation of individual SWNTs into surfactant micelles, and direct growth of individual SWNTs suspended in air between pillars. This paper compares the PL obtained from suspended SWNTs to published PL data obtained from encapsulated SWNTs. We find that emission peaks are blue-shifted by 28 meV on average for the suspended nanotubes as compared to the encapsulated nanotubes. Similarly, the resonant absorption peaks are blue-shifted on average by 16 meV. Both shifts depend weakly on the particular chirality and diameter of the SWNT.

With two thirds of single walled carbon nanotubes (SWNTs) predicted to be direct bandgap semiconductors [1,2], photoluminescence (PL) from the recombination of electron-hole pairs at the bandgap is to be expected. Although the SWNT material system has been studied for a decade, it is only recently that bandgap PL was reported, originally for solutions of purified SWNTs individually isolated inside surfactant micelles [3,4,5]. Upon illumination, infrared photoluminescence (PL) was detected from the nanotubes in solution, and it was determined that the light came from electron-hole recombination at the band edge. Shortly thereafter, we found that bare SWNTs suspended in air also emit bandgap PL [6]. In both cases the key is apparently to isolate the nanotubes, minimizing their interaction with the environment.

The potential for a material to luminesce obviously depends on its intrinsic band structure, but also other internal and external factors. Examples of internal factors include surface reconstructions, dislocations, dopants, and surface or bulk defects. Some external factors include the dielectric environment, electric fields, magnetic fields, and hydrostatic pressure, as well as any external chemical interactions. These factors may deplete or fill existing bands, or change the band structure entirely. As a result, PL emission energies may be shifted, or PL intensities may be altered, even to the point of destroying the luminescence altogether.

The case of the SWNT material system is particularly interesting, since for SWNTs, all the constituent atoms are on the surface, and they are therefore all exposed to the surroundings. Furthermore, it has already been clearly established that the electronic properties of a nanotube can be affected by its environment [7,8]. As an essentially direct optical probe of the electronic energy band structure, PL can provide useful, quantitative information about the importance of the interaction of nanotubes with their surroundings. This paper compares PL data obtained for SWNTs in two seemingly very different environments: in air, suspended between pillars on a substrate, and in encapsulating micelles, in aqueous solution.

## 1 Experimental details

In the first report of bandgap PL, the SWNTs were synthesized with the high-pressure carbon monoxide (HiPCO) process [9] and dispersed in a surfactant solution. This was centrifuged with the resulting supernatant being a solution of individually isolated SWNTs in soap micelles [3]. Subsequently, other surfactant solutions of SWNTs synthesized by laser vaporization and by arc discharge were found be active in PL [5]. In contrast, we isolated nanotubes in air, growing them directly by chemical vapor deposition (CVD) atop pillar arrays on patterned silicon substrates [6,10]. The pillars, nominally 180 nm in diameter, 300 nm high, and spaced by 400 nm were prepared on $SiO_2$ coated silicon substrates using synchrotron-radiation lithography. To catalyze SWNT growth, a thin layer of iron or cobalt (~1 nm) was evaporated in vacuum, covering all surfaces except the sides of the pillars.

Here, a pure methane CVD process was used to grow SWNTs [11,12]. Samples were heated to between 800ºC and 900ºC in a flow of argon (300 sccm at 66.5 kPa), after which argon was substituted for methane (300 sccm at 66.5 kPa) for one minute during which the nanotube growth occurs. After growth, the heater was turned off, the gas flow was stopped, the reactor was pumped out, and the samples were removed to air. It is very important to emphasize that the PL from these samples was measured "as-grown", without any post-growth processing or special handling.

A scanning electron microscope (SEM) image of SWNTs grown on pillars is shown in plan view in Fig. 1. The circular structures are pillars viewed from above, and suspended nanotubes are clearly seen extending from pillar to pillar, and from pillars down to the surface below. Because the catalyst deposition process was not selective, there are also many nanotubes lying on the bottom, flat surface. However, in SEM, suspended nanotubes generally show much better geometrical contrast than nanotubes on a surface. Therefore, in this case, only the suspended nanotubes are clearly visible. Photoluminescence is also selective, with luminescence detected from the suspended nanotubes, but not from the nanotubes on the flat surface. [6]

All PL spectra presented here were taken in air at room temperature. The PL was excited with one of two lasers, each in continuous wave mode, at ~1 mW excitation power. For fixed excitation wavelength, a frequency-doubled YAG laser (532 nm) was used. To create a PL excitation (PLE) map, a tunable titanium doped sapphire (Ti:sapphire) laser (725 to 837 nm) was used. In both cases an aspheric lens focused the excitation down to a 100 μm diameter spot, which on these samples corresponds to an area containing an ensemble of ~$10^4$ suspended nanotubes. Evidently the ensembles were relatively homogeneous across the substrate, because with such a large spot size the PL spectra were found to be largely position independent.

The luminescence was collected through the same lens used for excitation and it was dispersed by a single grating spectrometer (149 grooves/mm, 1250 nm blaze) onto a 512 element liquid nitrogen cooled InGaAs photodiode array, sensitive to wavelengths from the visible to ~1650 nm (0.75 eV). With the spectrometer centered at 1240 nm (1 eV), a spectrum covers the 900-1555 nm wavelength range (0.80-1.38 eV) with a resolution of about 1 nm (~1 meV). Spectra were typically integrated over a 30 s accumulation time. The relatively long integration time was chosen to reduce the signal-to-noise ratio, however the PL was sufficiently bright to be readily detected even using integration times of one second or less.

## 2 Results

A typical PL spectrum from an ensemble of SWNTs suspended between pillars is shown in Fig. 2, with excitation at 532 nm. Several peaks are seen, and in general the PL covers an extended infrared energy range. We have demonstrated previously that the PL comes from suspended nanotubes only [6]. No nanotube signal is detected when the laser excitation is focused onto an area where the nanotubes are lying directly on the substrate. The luminescence originates from electron-hole recombination at the band edge. The exact emission energy depends on the particular diameter and chirality of the SWNT.

As shown previously [4,5,6], the amplitude of PL peaks in a given spectrum depends strongly on the laser excitation energy. This is a consequence of the confinement of carriers normal to the nanotube axis. Around the circumference, the periodic boundary conditions give rise to a series of subbands. In such quasi-one-dimensional systems, the density of states as a function of energy has a series of sharp peaks, each of which is associated with a particular subband. Due to the high density of states, optical spectroscopy is very sensitive to allowed transitions between these sharp quasi-singularities. Tuning the laser excitation energy to resonate with such a singularity causes enhanced optical absorption, ultimately causing enhanced optical emission. As in the non-resonant case, the luminescence occurs after carriers relax and recombine at the lowest energy singularity.

To compare with previously published PLE maps for micelle encapsulated SWNTs [4,5], Figure 3a shows a color plot of the PL intensity as a function of emission energy and laser excitation energy for SWNTs suspended in air. The data is plotted on a linear color scale normalized to the highest peak. Several peaks can be clearly seen, each corresponding to a given SWNT species, with a specific diameter and chirality. The position of a given peak in emission energy ($x$-axis) is a measure of the energy of electron-hole pairs at the edge of the lowest subband ($E_{11}$), while the position in excitation energy ($y$-axis), is a measure of this energy at the edge of the second lowest subband ($E_{22}$). This resonance phenomenon in general, and the position of the peaks in particular, confirm that the luminescence does, in fact, originate from SWNTs.

To make the comparison clear, the same PLE map has been replotted in Figure 3b, but with the tabulated data of Ref. [4] (solid dots) and additional reference points (open circles) superimposed. Comparing the suspended nanotube peak positions to the micelle-encapsulated nanotube peak positions (solid dots), it is immediately obvious that they have different emission ($E_{11}$) and resonant excitation ($E_{22}$) energies. This offset for each individual peak is listed in Table 1. In all cases, both $E_{11}$ and $E_{22}$ for suspended nanotubes are blueshifted compared to micelle-encapsulated nanotubes. The average offset of $E_{11}$ is 28 meV with a standard deviation of 6 meV and the average offset of $E_{22}$ is 16 meV with a standard deviation of 8 meV. Adding the average offset (open circles) to the micelle-encapsulated nanotube peak positions produces a rather good match between the two data sets for all the peaks shown here. Even so, it is clear that, at least to some extent, the offset has some small diameter and/or chirality dependence. However, there is no obvious trend in this species dependence.

It is also worth noting that the $E_{22}/E_{11}$ ratio for these suspended nanotubes is in every case slightly lower than the micelle encapsulated nanotubes. For all the suspended nanotubes studied here, and for the corresponding species of micelle-encapsulated the average $E_{22}/E_{11}$ ratio is 1.7. As expected [4], this ratio varies substantially from species to species.

Fig. 3b also shows a dashed line that overlaps the faint streak visible in Fig. 3a. A straight line of unit slope fits very well, therefore the data can be assigned to a Raman process. Taking $E_{PL}$ as the detected energy and $E_L$ as the laser excitation energy, $E_{PL}$-$E_L$=0.71 eV (5700 cm$^{-1}$). This matches precisely with the expected position of the graphitic multiphonon 2D+2G Raman mode of 0.713 eV (5750 cm$^{-1}$), extrapolated from experimental visible-wavelength Raman data [13]. This and other phonon modes can be expected to have an impact on the relaxation of carriers from $E_{22}$ to $E_{11}$, and so play a role in the time dependence of nanotube PL. Such dynamical phenomena may well be responsible for enhancing the intensity of PL peaks that lie close to the phonon line.

**3 Discussion and conclusion**

We have demonstrated that suspended SWNTs and micelle-encapsulated SWNTs, seemingly in quite different physical environments, give rise to broadly similar optical spectra, with subtle, species-dependent differences. The emission and absorption energies of the seven species compared differed by 2 to 5%. The small size of this discrepancy suggests several possible conclusions. For one, optical emission and absorption may be robust properties of SWNTs, independent of the environment. This seems unlikely given recent findings that the choice of suspension can affect the luminescence [3,5]. Another possibility is that micelles may influence SWNT luminescence in the same way as suspension in air does. This seems too coincidental to be likely. A final, most likely possibility is that micelle encapsulation and pillar suspension in air may both provide sufficiently non-interactive environments that the PL is only weakly influenced.

There are several very promising directions for future research. This paper focused on comparing peak positions, but a more detailed comparison, including intensities, line shapes, dynamics and other features will be straightforward and useful. Importantly, it has been shown that the PL is at least somewhat environment dependent, even for these seemingly relatively mild environments. It will be interesting to explore just how much an influence the environment can have on the PL, and to elucidate the mechanisms by which the PL is affected, particularly as it relates to specific chiralities and diameters. With systematic study, it should be possible to use PL to probe the interaction of SWNTs with their environment, and, alternatively, it should be possible to prepare the SWNT environment in such a way as to tailor SWNT luminescent properties, as desired.

**Acknowledgements**. We thank R. L. Williams (NRC) for helpful suggestions regarding PL measurements, and D. Takagi (Meiji University) for a series of nanotube growths. This work was partially funded by the NEDO International Joint Research Grant Program.

**Table 1**

| Species[*] | Micelles[*] | | Pillars[§] | | Difference | |
|---|---|---|---|---|---|---|
| (n,m) | $E_{11}$(eV) | $E_{22}$(eV) | $E_{11}$(eV) | $E_{22}$(eV) | $\delta_{11}$(meV) | $\delta_{22}$(meV) |
| (12,1) | 1.059 | 1.556 | | | | |
| (11,3) | 1.036 | 1.565 | 1.060 | 1.593 | 24 | 28 |
| (10,5) | 0.992 | 1.577 | 1.011 | 1.601 | 19 | 24 |
| (9,7) | 0.937 | 1.569 | 0.964 | 1.588 | 27 | 19 |
| (10,6) | 0.898 | 1.640 | 0.929 | 1.651 | 31 | 11 |
| (9,8) | 0.877 | 1.533 | 0.904 | 1.550 | 27 | 17 |
| (13,3) | 0.828 | 1.631 | 0.865 | 1.640 | 37 | 9 |
| (12,5) | 0.829 | 1.559 | 0.860 | 1.566 | 31 | 7 |

[*] Taken from Ref. [4].

[§] Taken from this work.

**Figure 1.** Scanning electron micrograph of nanotubes on pillars. Suspended nanotubes are clearly visible.

**Figure 2**. Photoluminescence spectrum obtained from a large ensemble of suspended nanotubes. This spectrum was taken in air at room temperature with 1.0 mW laser excitation at 532 nm. The laser was focused to a 100 μm diameter spot, sampling ~$10^4$ pillars, and thus a comparable number of suspended SWNTs.

**Figure 3.** Photoluminescence excitation map. The color represents the luminescence intensity on a linear scale (see color bar). The PL intensity is plotted as a function of emission and excitation energies. The excitation was with a tunable continuous wave Ti:sapphire laser at 1.0 mW focussed to a 100 μm diameter spot. The spot sampled ~$10^4$ pillars and thus a comparable number of suspended SWNTs. The data are plotted without annotation in Figure 3(a). In Figure 3(b) the data are replotted, with annotation to facilitate comparison. The solid black dots show the locations of all peaks observed in Ref. [4] within these energy ranges. The open circles show the same data with a "best fit" constant offset of 28 meV in emission and 16 meV in excitation. The dashed line corresponds to a difference between excitation and emission energies of 0.71 eV, which overlaps the faint diagonal streak in the data, in good agreement with the expected position of the graphitic 2D+2G Raman mode.

**Table 1.** Peak positions of micelle-encapsulated nanotubes vs. suspended nanotubes. The emission energy ($E_{11}$) and resonant excitation energy ($E_{22}$) for SWNTs isolated in micelles (Ref. [4]) is compared to those for SWNTs suspended on pillars (this work).

The peak assignment to a particular nanotube species (n,m) is from Ref. [4]. The average difference is 28 meV in emission ($<\delta_{11}>$) and 16 meV in excitation ($<\delta_{22}>$). The average ratio of resonant excitation energy to emission energy ($<E_{22}/E_{11}>$) for the seven common peaks is 1.7 for both sets of data.

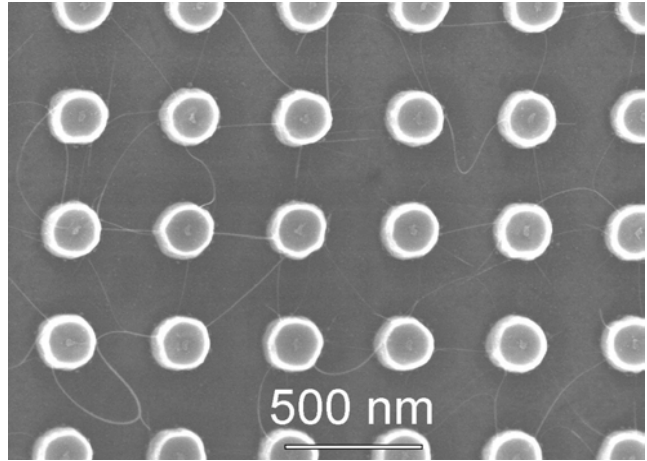

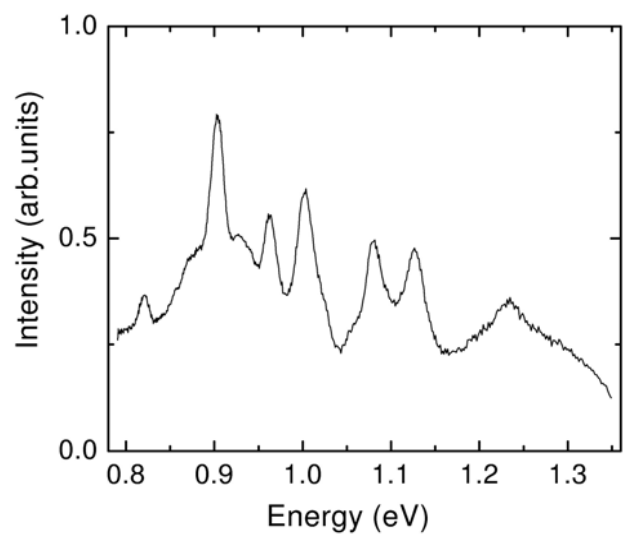

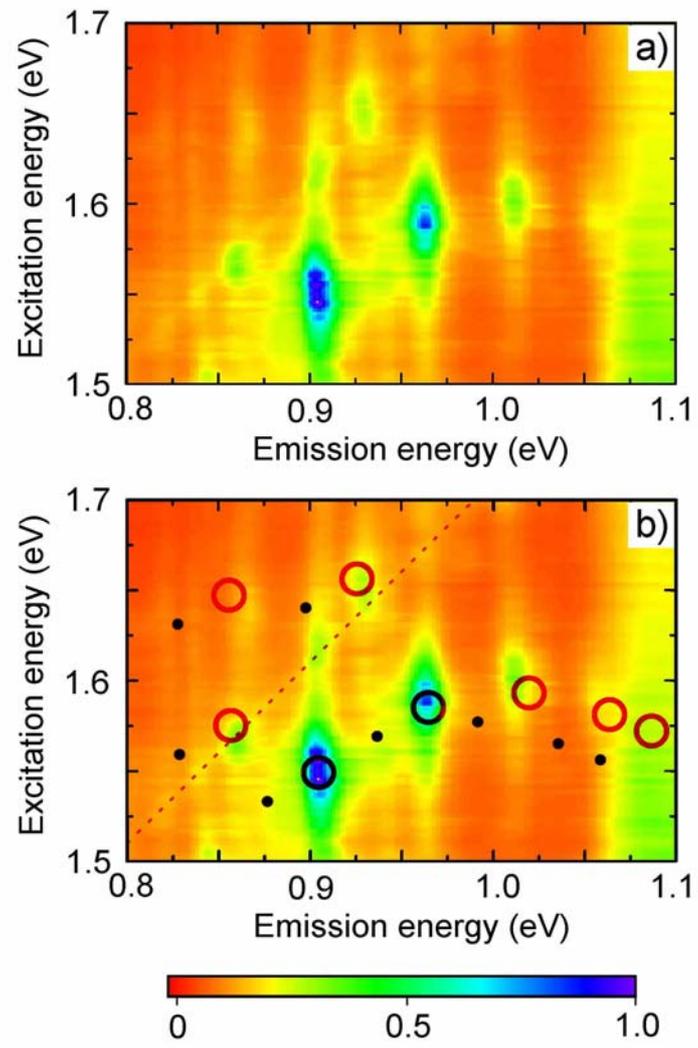